\documentclass[twocolumn]{aastex701}

\usepackage{graphicx}
\usepackage{subcaption}
\usepackage{amsmath}
\usepackage{amssymb}

\begin{document}

\title{Testing X-ray reprocessing through optical continuum reverberation mapping of the X-ray weak AGN Mrk 304}

\author[orcid=0000-0002-1380-1785]{Marcin Marculewicz}
\affiliation{Department of Physics and Astronomy, Wayne State University, 666 W. Hancock St, Detroit, MI, 48201, USA}
\email[show]{mmarculewicz@wayne.edu}  

\author[orcid=0000-0002-8294-9281]{Edward M. Cackett} 
\affiliation{Department of Physics and Astronomy, Wayne State University, 666 W. Hancock St, Detroit, MI, 48201, USA}
\email[]{ecackett@wayne.edu}  

\author{Russell W. Carroll}
\affiliation{Department of Physics and Astronomy, Wayne State University, 666 W. Hancock St, Detroit, MI, 48201, USA}
\email[]{rcarroll@miva.com}

\begin{abstract}
We investigate the origin of optical variability in the X-ray weak active galactic nucleus (AGN) Mrk~304, in order to test the viability of the standard X-ray reprocessing model. We present the results of a multi-wavelength monitoring campaign, including simultaneous $u, g, r, i, z$ photometry from the Zowada Observatory and archival \textit{NuSTAR} X-ray data. Using time-series analysis with both \texttt{PyCCF} and \texttt{JAVELIN}, we measure inter-band time lags that increase with wavelength.
While the lag-wavelength spectrum is broadly consistent with the standard thin disk model ($\tau \propto \lambda^{4/3}$), we identify a significant excess lag in the $i$-band, which we attribute to contamination from the H$\alpha$ emission line, and a similar excess in the $u$-band. 
We confirm that the source is energetically X-ray weak, estimating a broad-band (0.03--200 keV) X-ray to variable optical/UV disk luminosity ratio of $L_{\text{X}} / L_{\text{disk}} \approx 0.19$. This low ratio demonstrates that the X-ray corona is energetically insufficient to power the observed optical/UV variability through reprocessing alone. 
Frequency-resolved lag analysis reveals that the observed lags are strongly timescale-dependent, becoming consistent with zero at frequencies higher than 0.03 day$^{-1}$ (i.e., timescales shorter than approximately 30 days). This suggests that the reprocessing region extends beyond 30 light-days.
Taken together, our results collectively challenge the standard X-ray disk-reprocessing model for this object, suggesting either EUV irradiation of the BLR/wind producing diffuse continuum emission, or a different origin for the reverberation signal.
\end{abstract}


\section{Introduction}
The small angular size on the sky of the inner regions of active galactic nuclei (AGN) limits observational tests of the structure and geometry close to the supermassive black hole. The reverberation mapping (RM) technique \citep[e.g.,][]{Blandford82, Peterson14, Cackett21RMreview} measures time delays, or lags, between light curves at different wavelengths.  The assumption is that the lag originates because of differences in path length between light that comes directly to us compared to light from other regions in the AGN, where it has been absorbed and re-emitted.  Correlated UV/optical continuum light curves have been observed for many years \citep[e.g.,][]{peterson91,edelson96,collier98,sergeev05}. These are typically interpreted as due to reverberation from the accretion disk \citep[AD;][]{collin-souffin91, Krolik91, clavel92, collier99, Cackett07}. In this scenario, emission from the X-ray corona irradiates the surrounding AD, transferring a significant amount of its energy to the disk's surface. The AD absorbs a significant portion of the high-energy radiation. It then reprocesses this radiation, thermally re-radiating the energy primarily in the ultraviolet and optical wavelength regimes. Crucially, this reverberation - the time-delayed response of the disk to variations in coronal X-ray emission - encodes information about the physical separation between the corona and the disk. Through measuring lags, $\tau$, between continuum light curves at different wavelengths, we can test the geometry, with lags expected to follow $\tau \propto \lambda^{4/3}$ for an optically thick, geometrically thin disk \citep[e.g.,][]{collier98, Cackett07}.

For X-rays to drive thermal reprocessing at longer wavelengths, there must be sufficient energy in the variable X-rays  \citep[e.g.,][]{edelson96}. It was realized early that AGN UV/optical variability cannot be fully explained by X-ray reprocessing of the AD alone \citep[e.g.,][]{clavel92}. In a particularly clear example, \cite{Uttley2003} demonstrated that the long timescale ($>$year) UV variability amplitude in NGC~5548 substantially exceeded the X-ray variability amplitude, requiring the need for there to be intrinsic disk variability not driven by reprocessing. This idea was supported by \cite{Breedt2009}, who conducted concurrent X-ray and optical monitoring of Mrk~79 over a period of more than five years. Their observations revealed a characteristic two-component variability structure: on short timescales of days to tens of days, optical and X-ray variations were well correlated and consistent with disk reprocessing. However, on longer timescales, the optical light curve evolved independently of the X-rays. More recently, the measurement of long timescale optical variations leading the UV \citep{JVHS_F9_2020,Yao2023} is indicative of intrinsic disk fluctuations propagating inward.  This is supported by three-dimensional magnetohydrodynamic simulations that demonstrate that AD turbulence can produce the observed UV variability with minimal contribution from X-ray irradiation \citep{Secunda2024}, but that X-ray reprocessing does lead to lags if there is sufficient X-ray luminosity and low albedo \citep{Secunda2025}.

Recent intensive, multi-wavelength RM campaigns using Swift and NICER have revealed further challenges to the standard X-ray reprocessing model \citep{Edelson2015, Edelson19, gardner17, partington23, Partington2024}. While X-ray to UV correlations are excellent in some cases \citep{McHardy2018, Partington2025}, other observations reveal poor correlations, and sometimes no correlation at all, contradicting the prediction of strongly correlated light curves.  Moderately poor X-ray/UV correlations can be explained by a dynamic X-ray corona \citep{panagiotou22}. However, careful consideration of changes in the spectral energy distribution requires more than just reverberation from the AD \citep{Hagen2024,Hagen2025}.

There are additional challenges to the interpretation of simple X-ray disk reprocessing. For instance, these intensive RM campaigns have also commonly observed that the lag in the $u$ band is in excess of the extrapolation of the lags in the other bands \citep[e.g.,][]{Edelson2015,Edelson19,Fausnaugh2016RM_NGC5548,McHardy2018, JVHS_F9_2020}. In NGC 4593, spectroscopic monitoring with HST revealed this $u$-band excess to be a broad bump and discontinuity around the Balmer jump \citep{Cackett18}. This is consistent with the picture where diffuse continuum emission from the Broad Line Region (BLR) itself is contributing significantly to the lags. Such diffuse continuum emission should be present at all wavelengths and can mimic the $\tau \propto \lambda^{4/3}$ dependence expected for a disk \citep{KoristaGoad2001, Korista2019, Lawther2018, Netzer22}. Since the sizescale of the BLR is larger than the disk, the timescale on which the reverberation occurs should be longer.  Recent analysis using either time-domain techniques, such as detrending \citep{McHardy2018,pahari20,Vincentelli2021,Cackett2023_Mrk817, MM2024,MM26} or frequency-resolved techniques \citep{Cackett2022_freq_rs, Lewin2023, Lewin2024, lewin2025, Drewes2026} demonstrate that lags are longest at the lowest frequencies (longest timescales), and shortest at the highest frequencies (shortest timescales). The low-frequency lags are longer than expected from a disk alone, suggestive of a BLR diffuse continuum.

The existence of X-ray weak AGN poses a further challenge to the standard X-ray reprocessing scenario. X-ray weak sources are characterized by X-ray emission that is significantly suppressed below the expected correlation with ultraviolet (UV) luminosity. This is quantified by deviations from the standard $\alpha_{\rm OX}$--$L_{\rm UV}$ relation, where $\alpha_{\rm OX}$ is the optical-to-X-ray spectral slope \citep{Just2007}. These sources exhibit X-ray deficits that deviate markedly from the empirical correlation defined for normal AGN populations. This raises a key question: if the X-ray corona is the main source of dissipated accretion power and drives the accretion flow, how can such a significant X-ray deficit occur while the AD continues to emit substantial UV/optical radiation? This apparent decoupling challenges our understanding of what drives the variability.

In Section~\ref{sec:data}, we present the data used, Section~\ref{sec:timeser} describes our time lag analysis, and in Section~\ref{sec:disc} we discuss the implications of our results on the standard X-ray reprocessing scenario.

\section{Data}\label{sec:data}

\subsection{UV/Optical}

\begin{deluxetable}{rllllll}
\centering
\tablecaption{Optical light curve properties of Mrk 304 from the Zowada Observatory.\label{tab:lc_properties}}
\tabletypesize{\footnotesize}
\tablehead{
\colhead{Band} & \colhead{N} & \colhead{Start} & \colhead{End} & \colhead{Length} & \colhead{$\langle \Delta t \rangle$} & \colhead{$F_{\rm var}$}\\
&&&&[d]&[d]&}
\startdata
u & 109 & 2025-06-14 & 2025-11-28 & 167 & 1.5 & 0.10 \\
g & 131 & 2025-06-14 & 2025-11-30 & 169 & 1.3 & 0.08 \\
r & 126 & 2025-06-14 & 2025-11-30 & 169 & 1.4 & 0.07 \\
i & 125 & 2025-06-14 & 2025-11-30 & 169 & 1.4 & 0.05 \\
z & 129 & 2025-06-14 & 2025-11-30 & 169 & 1.3 & 0.05 
\enddata
\tablecomments{$N$ is the number of epochs of observations, $\Delta t$ is the cadence between consecutive observations, and $F_{\rm var}$ is the fractional variability amplitude \citep{Vaughan2003}.}
\end{deluxetable}

During the monitoring of Mrk 304, 300s exposures were taken in each of the filters.  Multiple exposures were taken in each filter, with 4 exposures per epoch in $u$ and $z$, and 2 exposures per epoch in $g$, $r$, and $i$.  The images were reduced following standard bias and dark subtraction and flat-fielding procedures.  Aperture photometry was performed on the AGN and comparison stars using a circular source aperture with a 5-pixel radius, and an annular background aperture with inner and outer radii of 20 and 30 pixels, respectively.  3 comparison stars were used in most bands, but for the $u$ band, 2 different comparison stars were used, given the fewer appropriate stars available. Relative photometry was performed under the assumption that the sum of the comparison stars remains constant from epoch to epoch.  Photometry from exposures taken on the same night was averaged together. Epochs where the comparison star fluxes deviate significantly from a constant were rejected.

\begin{figure*}
\centering
\includegraphics[width=1\textwidth]{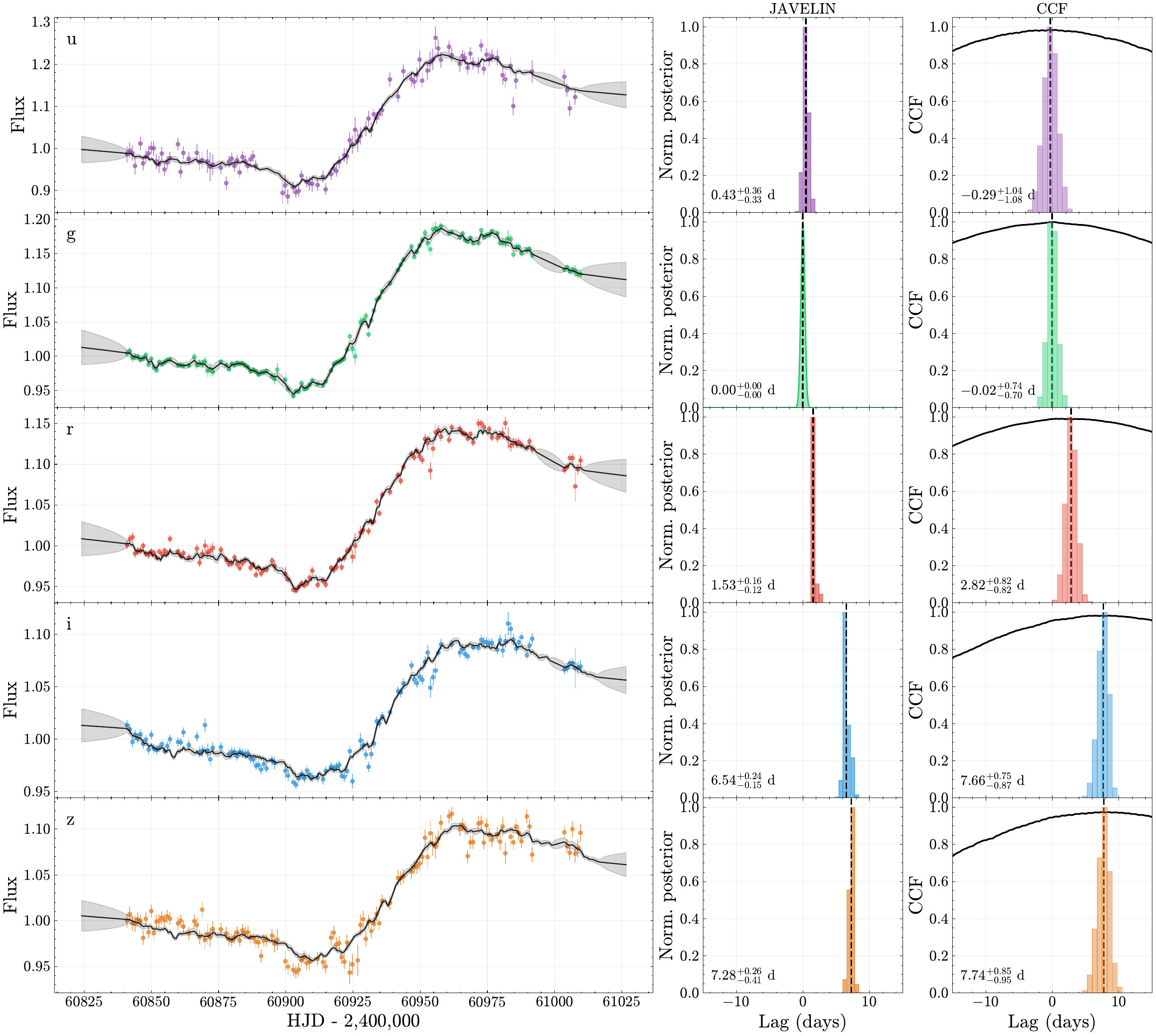}
\caption{{\it Left:} Zowada light curves of Mrk 304 during the 2025 monitoring, from the shortest wavelength ($u$, top) to the longest wavelength ($z$, bottom). The solid line and shaded area show the \texttt{JAVELIN} model and uncertainty. {\it Middle:} histograms of the probability distribution of lag measurements from \texttt{JAVELIN}. {\it Right:}, histograms of the centroid time-delay distributions from \texttt{PyCCF}. The solid black line shows the CCF between each wave band and the $g$ light curve.}
\label{fig:lightcurves_ccf}
\end{figure*}

To flux calibrate the light curves, we measure the instrumental magnitude of all stars in the field during a reference epoch.  We fit the instrumental magnitudes to the catalog magnitudes to determine the zeropoint. For $g$, $r$, $i$ and $z$ we use the APASS DR10 catalog \citep{henden2018} as reference. This catalog does not contain $u$ magnitudes in this particular field, and so we instead rely on the SDSS DR12 catalog for the $u$ band. The resulting light curves are shown in Figure~\ref{fig:lightcurves_ccf}, with 131 epochs over 169 days in the $g$ band.  The light curves are highly variable, with a variability amplitude of $F_{\rm var} = 8$\% in $g$.

\subsection{ZTF}
We investigate Mrk 304 in two bands ($g$ and $r$)
using ZTF data. We retrieve the ZTF $g$- and $r$-band light
curves via the IRSA GATOR ZTF
catalog service\footnote{https://irsa.ipac.caltech.edu/applications/Gator/}. ZTF is a time-domain survey with the 1.2 m
Schmidt telescope at the Palomar Observatory \citep{ZTF_data_processing2019}. The available data are from 2018 May 11 until 2025 October 21. The cadence $\langle \Delta t \rangle$ varies between seasons from daily 0.95 up to 4.03 days. For details of the observations, see Appendix Tab. \ref{tab:season_properties}. 

\subsection{X-ray}

We analyze archival \textit{NuSTAR} observations of Mrk 304 (ObsID: 60160785002; 2024 November 07), providing the first hard X-ray measurements since \cite{Brinkmann2004}. The past XMM-Newton observations of the Seyfert~1 galaxy Mrk~304 reveal a complex broadband X-ray spectrum dominated by a power-law continuum with photon index $\Gamma \sim 1.9$ and a 2--10~keV luminosity of $\sim 7 \times 10^{43}$\,erg\,s$^{-1}$ \citep{Brinkmann2004}.

We reduced the 2024 data using \texttt{NuSTARDAS} v2.1.4 within \texttt{HEASoft} v6.34, and the most recent \texttt{CALDB} available at the time of analysis. Cleaned event files for both focal plane modules (\texttt{FPMA} and \texttt{FPMB}) were produced with \texttt{nupipeline}. Source and background spectra were then extracted with \texttt{nuproducts} using circular regions: a 72\arcsec\ diameter source region centered on the target and a 60\arcsec\ diameter background region placed in a source-free area on the same detector chip. The spectra were grouped with \texttt{ftgrouppha} using optimal binning (\texttt{optmin}) with a minimum of 25 counts per bin.

\begin{figure}
\centering
\includegraphics[width=1\linewidth]{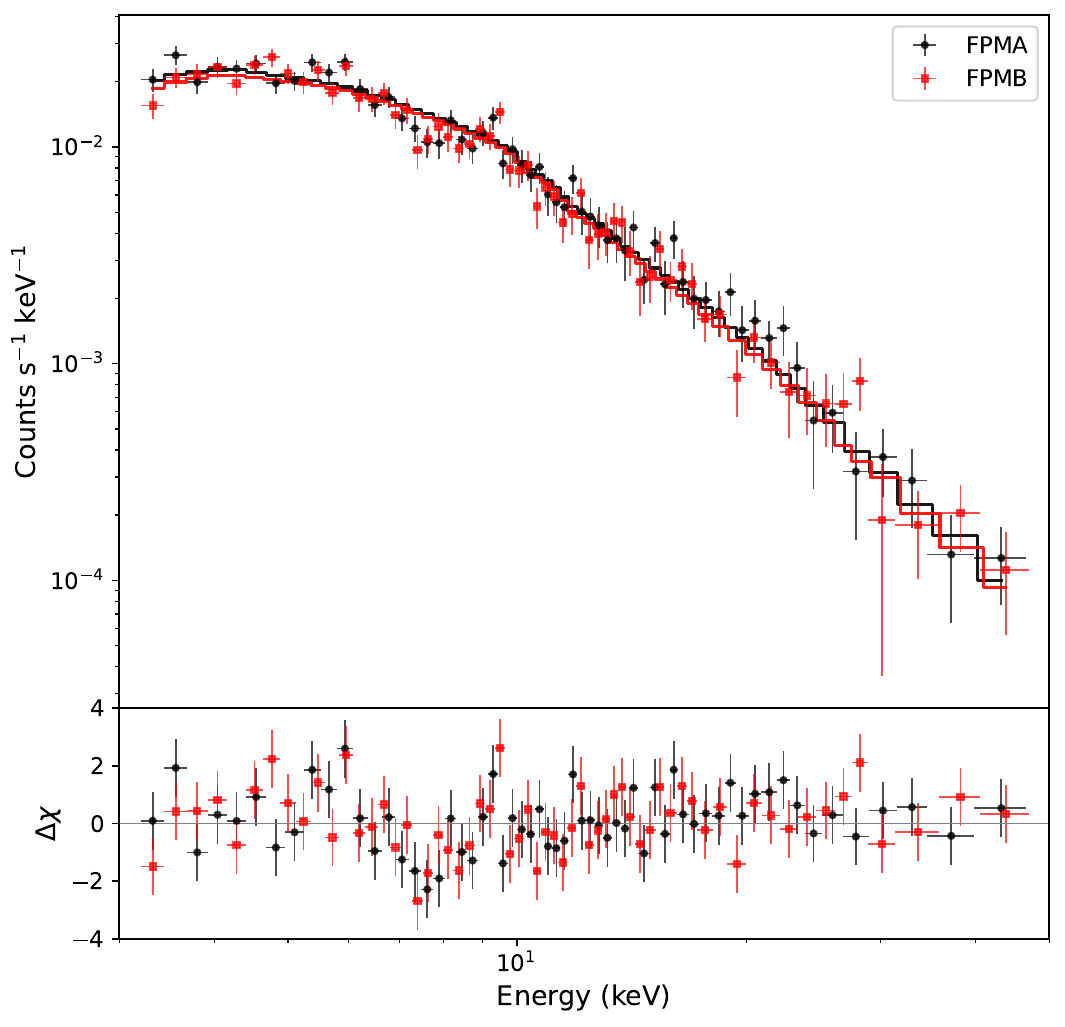}
\caption{NuSTAR spectrum of Mrk 304 from 2024 November (\texttt{FPMA} in black, and \texttt{FPMB} in red). Solid lines show the best-fitting model. Residuals to the best-fitting model are presented in the lower panel.}
\label{fig:nustar_fit}
\end{figure}

We fitted the \texttt{FPMA} and \texttt{FPMB} simultaneously in \textsc{XSPEC} over the 3--50~keV range, excluding higher energy bins where the source becomes background dominated.  The spectra are well described by an absorbed power-law model, \texttt{constant*TBabs*powerlaw}, where the Galactic absorption was fixed at $N_{\rm H}=0.05\times10^{22}\,\mathrm{cm^{-2}}$ and the constant allows for a small offset in flux between the FPMs. The NuSTAR spectrum and best-fitting model are shown in Fig. \ref{fig:nustar_fit}.

The fit is statistically acceptable, with $\chi^{2}/{\rm d.o.f.}=129.2/118$, and yields a photon index of $\Gamma = 1.8 \pm 0.1$ and an absorbed 2--10~keV flux of $F_{2-10}=(6.9\pm0.2 )\times10^{-12}\,\mathrm{erg\,cm^{-2}\,s^{-1}}$. At a luminosity distance of 301~Mpc, this corresponds to $L_{2-10}=(7.6\pm0.3)\times10^{43}\,\mathrm{erg\,s^{-1}}$. This is broadly consistent with the previous X-ray luminosity measured by \cite{Brinkmann2004}.

\section{Light curve analysis}\label{sec:timeser}

\subsection{Lag Measurements}
We measure the time lags between the light curves using two complementary approaches: the interpolated cross-correlation function (ICCF) \citep{{Peterson04}} using \texttt{PyCCF} \citep{PyCCF_MSun_2018}, and \texttt{JAVELIN} \citep{Zu2013jav}. \texttt{PyCCF}, calculates the cross-correlation function (CCF) between two light curves. Since the data are unevenly sampled to allow the CCF to be calculated at any time shift, this method linearly interpolates between the data points.  To estimate the lag and uncertainties, \texttt{PyCCF} applies iterative randomization, which involves perturbing flux values and sub-sampling data \citep{Peterson1998PASP} to create different realizations of the light curves.  The CCF is determined for each realization, and the peak and centroid of the CCF are measured. The centroid is calculated using values that exceed 80\% of the maximum correlation coefficient (setting $r_\mathrm{thresh} = 0.8$ for the Zowada data).  This is repeated for 10,000 realizations, creating a centroid and peak lag distribution. The reported lags are determined as the median of the centroid and peak distributions, and the uncertainties are determined using the 16th and 84th percentiles.

The second approach uses a parametric variability model, specifically, assuming AGN variability is described by a damped random walk (DRW) to more realistically interpolate between data points. \texttt{JAVELIN} adopts this strategy by fitting a DRW to a reference light curve and deriving lags by smoothing the multi-band data as it deviates from this template. \texttt{JAVELIN} produces smaller uncertainties because it assumes a parametric variability model. By contrast, \texttt{PyCCF} makes no such assumptions, resulting in larger uncertainties being reported \citep{Yu2020_javunc}. Lags from both methods are given in Tab. \ref{tab:ccf_jav}. Both methods show statistical agreement. The best-fitting \texttt{JAVELIN} model, lag distributions, and the CCF are shown along with the light curves in Fig. \ref{fig:lightcurves_ccf}.
During the whole analysis, we use the $g$-band as the reference band, as it has the highest number of epochs and best cadence, see Tab. \ref{tab:lc_properties}.

The CCF is determined for each realisation, from which the peak and centroid are measured. Notably, the maximum correlation coefficients ($r_{\mathrm{max}}$) between the different wavelength light curves are very high, approaching unity across all bands despite the low X-ray luminosity. These values are consistent with, or comparable to, the ($r_{\mathrm{max}}$) values reported for other AGN in \citet[][their Fig. 2]{Edelson19}.

\begin{deluxetable}{cccc}
\centering
\tablecaption{Lags (in days) from \texttt{PyCCF} (centroid and peak) and \texttt{JAVELIN}.\label{tab:ccf_jav}}
\tablehead{\colhead{Band} & \colhead{CCF Centroid} & \colhead{CCF Peak} & \colhead{JAVELIN}}
\startdata
$g$ & $-0.0^{+0.7}_{-0.7}$ & $-0.1^{+0.4}_{-0.2}$ & $0.0^{+0.0}_{-0.0}$ \\
$u$ & $-0.3^{+1.0}_{-1.1}$ & $-0.4^{+1.0}_{-1.8}$ & $0.4^{+0.4}_{-0.3}$ \\
$r$ & $2.8^{+0.8}_{-0.8}$ & $1.6^{+2.0}_{-1.0}$ & $1.5^{+0.2}_{-0.1}$ \\
$i$ & $7.7^{+0.8}_{-0.9}$ & $7.3^{+1.3}_{-1.7}$ & $6.5^{+0.2}_{-0.2}$ \\
$z$ & $7.7^{+0.9}_{-1.0}$ & $7.7^{+1.9}_{-1.1}$ & $7.3^{+0.3}_{-0.4}$ \\
\enddata
\end{deluxetable}

\subsection{Lag–Wavelength Relation}
\begin{figure}
\centering
\includegraphics[width=1\linewidth]{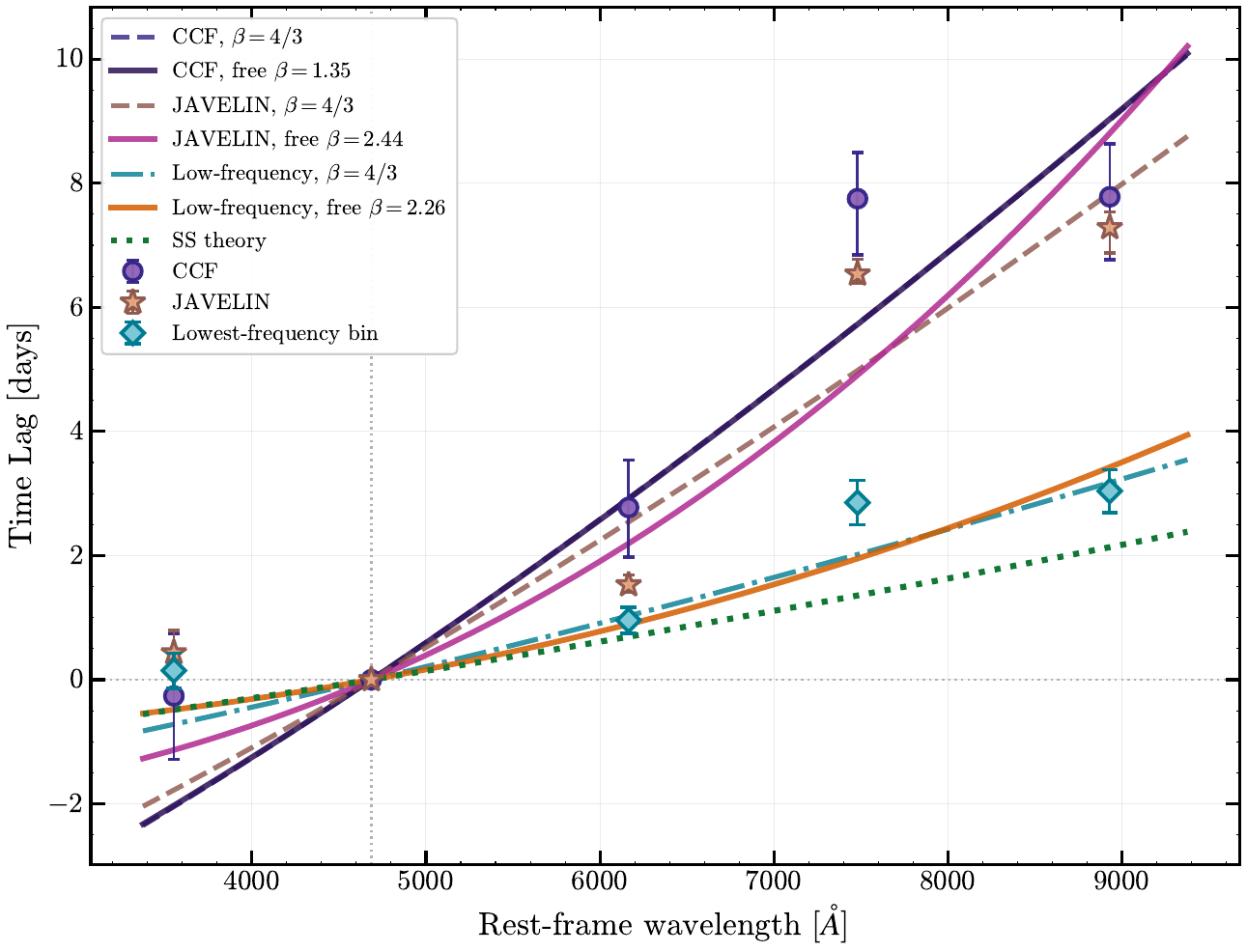}
\caption{Lag vs. wavelength measured with \texttt{JAVELIN} (copper) and \texttt{PyCCF} (purple). Lags are shown in the observed frame and are calculated with respect to the $g$ band (vertical dashed line at 4770 \AA). 
Light blue diamonds are lags from the lowest frequency bins for each band from the frequency-resolved analysis.
The green line is SSD prediction assuming $\dot{m} = 0.1$ an average from Tab. \ref{tab:bolometric_estimates}, with $M_{\text{BH}} = 10^{8.3} \, M_{\odot}$ \citep{Watabe2009}.}
\label{fig:lag_wavelength}
\end{figure}

We show lag plotted as a function of wavelength in Fig.~\ref{fig:lag_wavelength}.
The standard thin accretion disk model \citep[SSD,][]{SS73} predicts that lag scales with wavelength as $\tau \propto \lambda^{4/3}$. To test this relation against the observed lags, we fit:

\begin{equation}
\label{eq:tau}
\tau = \tau_0 \left[\left(\frac{\lambda}{\lambda_0}\right)^{\beta} - 1\right]
\end{equation}

where $\tau $ is the time lag, $\tau_0$ is a normalization constant, $\lambda_0 = 4770$ \AA\ (the $g$ band reference wavelength), and $\beta$ is the power-law index. We consider two scenarios: $\beta = 4/3$ (standard thin disk) and $\beta$ as a free parameter to determine which model best describes the data. We find that both agree with the standard thin disk $\beta = 4/3$.  We obtain better $\chi^2$ values for fits to the CCF lags, as expected due to the much smaller lag uncertainties we obtain with \texttt{JAVELIN}. The $\chi^2_\nu$ for \texttt{PyCCF} is 3.65.
Fit parameters are given in Tab. \ref{tab:disk_lag_fit}.

\begin{deluxetable}{lcccc}
\centering
\tablecaption{Disk lag fit parameters using CCF, \texttt{JAVELIN}, and frequency-resolved (FR) lags.\label{tab:disk_lag_fit}}
\tablehead{
\colhead{Dataset} & \colhead{$\beta \pm \sigma_\beta$} & \colhead{$\tau_0 \pm \sigma_{\tau_0}$ [days]} & \colhead{$\chi^2$} & \colhead{$\chi^2_\nu$} }
\startdata
CCF fixed & $4/3$ & $6.7 \pm 0.5$ & 10.9 & 3.7 \\
JAV fixed & $4/3$ & $5.8 \pm 0.2$ & 159.8 & 53.3 \\
CCF free & $1.4 \pm 0.8$ & $6.5 \pm 5.4$ & 10.9 & 5.5 \\
JAV free & $2.4 \pm 0.3$ & $2.3 \pm 0.4$ & 134.0 & 67.0\\
\hline
FR fixed & $4/3$ & $2.3 \pm 0.2$ & 15.2 & 5.1 \\
FR free & $2.3 \pm 0.8$ & $1.1 \pm 0.6$ & 12.7 & 6.3\\
\enddata
\end{deluxetable}

With both \texttt{PyCCF} and \texttt{JAVELIN}, we are able to ensure the robustness of our result using independent techniques, enabling cross-verification, and the existence of $i$ and $u$ bands excess --  both the $u$ and $i$ band lags sit above the best-fitting relationship in Fig.~\ref{fig:lag_wavelength}. Mrk 304 is not the only source that a $u$ band excess is observed, \citep[e.g.,][]{Edelson2015,Fausnaugh2016RM_NGC5548,Cackett18,Miller2025,MM26}. The excess emission observed in the $u$ band may be associated with diffuse continuum emission from the BLR. Photoionised gas in the BLR can produce a diffuse continuum through processes such as free-bound and free-free emission. This can contribute significantly to the observed broadband flux and lead to differences in the intrinsic accretion disk continuum \citep[e.g.,][]{KoristaGoad2001, Korista2019}. The $u$-band excess, in particular, is consistent with the expected contribution from the Balmer continuum and blended emission lines. 

\begin{figure}
\centering
\includegraphics[width=1\linewidth]{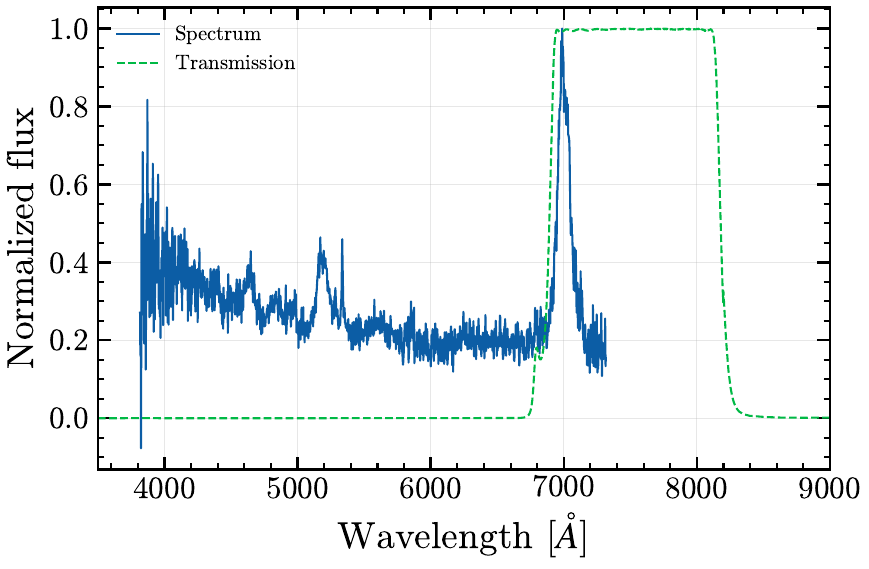}
\caption{ Observed spectrum of Mrk 304 in blue with transmission curve of the $i$-band (green dashed line).}
\label{fig:spectrum}
\end{figure}

The $i$-band excess, however, likely arises from a strong and variable H$\alpha$ line, which can increase the lag in that specific band \citep[e.g.,][]{Miller2023}. As the spectrum\footnote{\url{https://buil.astrosurf.com/galaxies/spectra.html}} of Mrk 304 shows (Fig.~\ref{fig:spectrum}), at the redshift of $z = 0.066$, the H$\alpha$ line falls into the $i$ band filter.

Given the AGN parameters derived for Mrk 304, the SSD model predicts a $\tau_0$ value of approximately 1.74 days according to Eq. 12 from \cite{Fausnaugh2016RM_NGC5548} assuming standard parameters: $\eta = 0.1$, $f_{\rm col} = 1.0$, $X = 2.49$, and $\lambda_0 = 4770$\,\AA. However, our best-fitting model yields a value of approximately $6.7 \pm 0.5$ days, which is significantly larger than the SSD prediction. This discrepancy between the predicted and observed timescale is consistent with recent reverberation mapping studies, in which $\tau_0$ values frequently exceed the simplest SSD expectations \citep[e.g.,][]{Guo2022_ZTF_arxiv, Steyn26}, though we note that more advanced models can reproduce the disk sizes but usually require a higher X-ray source \citep{Kammoun2021, Kammoun2023}.

\subsection{ZTF lags}
\begin{figure}
\centering
\includegraphics[width=\linewidth]{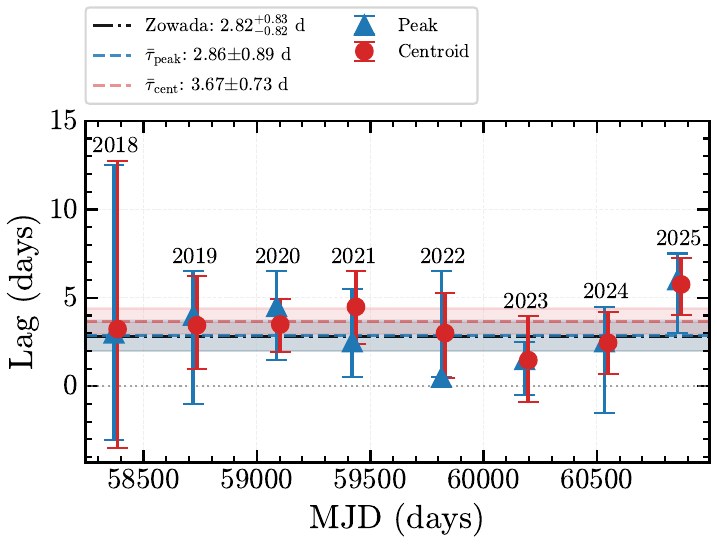}
\caption{Centroid (red) and peak (blue) $g-r$ lag measurements from ZTF. Each point corresponds to a separate year from 2018 to 2025. Red and blue horizontal dashed lines and shaded regions indicate the mean and uncertainty from the ZTF measurements. The black horizontal dashed line and shaded region indicate the centroid lag measurement from Zowada.}
\label{fig:ztf_lags}
\end{figure}

We measure lags from the ZTF light curves during each observing season to avoid seasonal gaps that may introduce spurious measurements. The results are summarised in Fig.~\ref{fig:ztf_lags} and Table~\ref{tab:ccf_mrk304}. Lags are recovered in all eight seasons spanning 2018--2025, with generally good consistency across years.

The CCF centroid threshold was set to $r_{\mathrm{thresh}} = 0.9$. We tested lower thresholds of 0.5 and 0.8, but found that in certain seasons -- most notably 2020 and 2023 -- the broad CCF profile caused the extended wings of the cross-correlation function to skew the centroid measurements. In the case of 2020 in particular, the light curves exhibit a near-monotonic long-term trend during that season, which produces a broad, flat CCF peak without a well-defined maximum. Under these conditions, including the low-correlation wings in the centroid calculation biases the lag estimate, and adopting $r_{\mathrm{thresh}} = 0.9$ mitigates this by restricting the centroid to the core of the peak.

The 2018 season carries the largest uncertainties, likely reflecting the significantly lower peak correlation coefficient ($r_{\mathrm{max}} \sim 0.4$) compared to other years, which broadens the FR/RSS bootstrap distributions. The remaining seasons yield well-constrained centroid lags broadly consistent with one another within their respective uncertainties, with the exception of 2025, which shows a somewhat longer centroid lag of $5.8^{+1.5}_{-1.7}$\,days. 

\begin{deluxetable}{lcc}
\centering
\tablecaption{CCF $g$--$r$ lags (in days) for Mrk304 from ZTF light curves. \label{tab:ccf_mrk304}}
\tablehead{\colhead{Year} & \colhead{CCF Centroid} & \colhead{CCF Peak}}
\startdata
2018 & $3.2^{+9.5}_{-6.7}$ & $3.0^{+9.5}_{-6.0}$ \\
2019 & $3.4^{+2.8}_{-2.5}$ & $4.0^{+2.5}_{-5.0}$ \\
2020 & $3.5^{+1.5}_{-1.6}$ & $4.5^{+2.0}_{-3.0}$ \\
2021 & $4.5^{+2.0}_{-2.1}$ & $2.5^{+3.0}_{-2.0}$ \\
2022 & $3.0^{+2.3}_{-2.5}$ & $0.5^{+6.0}_{-0.0}$ \\
2023 & $1.5^{+2.5}_{-2.4}$ & $1.5^{+1.0}_{-2.0}$ \\
2024 & $2.5^{+1.8}_{-1.8}$ & $2.5^{+2.0}_{-4.0}$ \\
2025 & $5.8^{+1.5}_{-1.7}$ & $6.0^{+1.5}_{-3.0}$ \\
\enddata
\end{deluxetable}

\subsection{Frequency-resolved analysis}

As described in the Introduction, studies utilizing lag analysis that goes beyond measuring just an average lag have revealed timescale-dependent lags.  Analysis that involves detrending is sensitive to the chosen detrending method, and so instead we choose to apply a frequency-resolved approach  \citep{uttley2014, Cackett2022_freq_rs}. We do this with two different methods. Firstly, since the cadence of the monitoring is consistently close to 1-day throughout the campaign, we linearly interpolate the light curves onto a regular 1-day time grid.  We are then able to directly calculate the lags by taking the fast Fourier transform of the light curves to calculate the cross spectrum and the coherence \citep{uttley2014}. We bin the cross spectrum in bins of equal size in logarithmic frequency, and combine the two lowest frequency bins together given the few Fourier frequencies there.  Secondly, we also use the STELA toolkit\footnote{https://github.com/collinlewin/STELA-Toolkit}, which employs Gaussian processes to interpolate the light curves and calculate the lags \citep{Lewin2023,Lewin2024,lewin2025}. The lags recovered using the STELA toolkit are consistent with what we recover with the simple linear interpolation approach, though the lowest frequency bin in the $i$ and $z$ bands is slightly shorter using STELA.  We show the lags as a function of frequency in Fig.~\ref{fig:lagfreq}. A direct comparison of the STELA lags to the linearly interpolated lags is shown in the Appendix, in Fig. \ref{fig:linear_vs_stela}.  The Appendix also provides details of the kernel choice for the STELA analysis.

Comparing the frequency-resolved lags with the CCF centroid lag, we see that for the $r$, $i$ and $z$ bands the CCF centroid lag is only consistent with the frequency-resolved lags at the very lowest unbinned Fourier frequency, while the lowest binned frequency is substantially shorter than the CCF value. We show the lags vs wavelength for the lowest frequency bin in Fig.~\ref{fig:lag_wavelength}. Fitting Equation \ref{eq:tau}, we find the best-fitting relation is close to the SSD prediction for $\beta$ = 4/3, which yields $\tau_0 = 2.3 \pm 0.2$ days.

Another particularly important result to note is that the lags are consistent with zero at frequencies higher than approximately 0.03 day$^{-1}$, corresponding to timescales shorter than about 30 days. This indicates that the region responsible for the reprocessing must be larger than about 30 light-days.  Therefore, the frequency-resolved analysis demonstrates that the lags are only present in the longest timescales probed by the light curves.

\begin{figure*}
\centering
\includegraphics[width=1.0\linewidth]{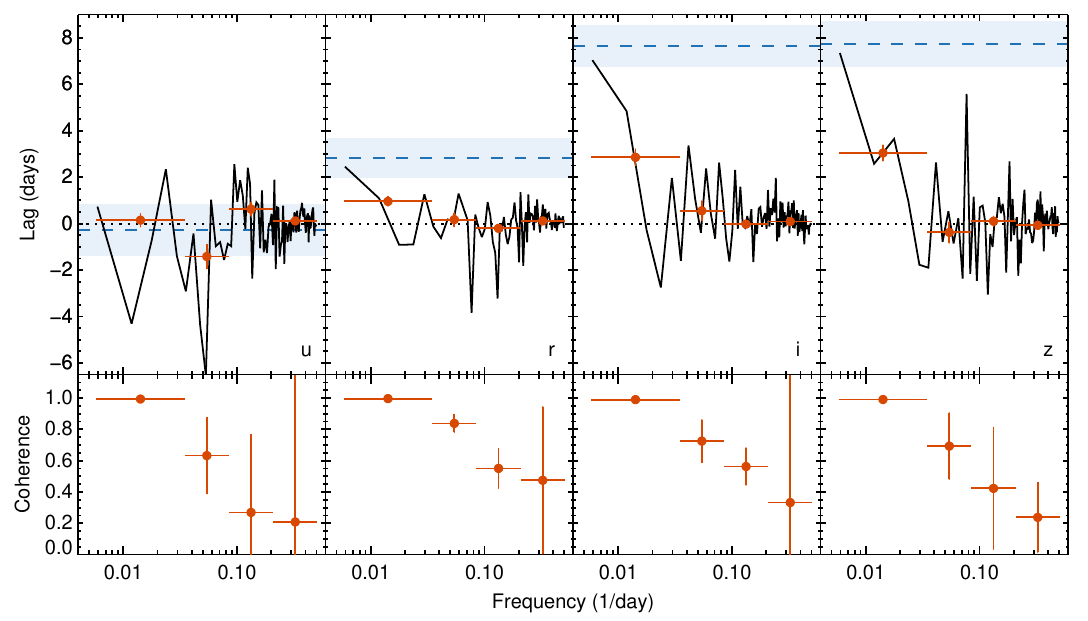}
\caption{The top panels show lag with respect to the $g$ band, as a function of frequency for each of the $u, r, i$ and $z$ light curves, calculated linearly interpolating the light curves onto a regular 1-day grid.  The black lines show the unbinned lags, while the orange circles show the binned lags. The blue dashed line and region show the CCF centroid lags. The bottom panel shows the coherence for each band with respect to $g$.}
\label{fig:lagfreq}
\end{figure*}
    
\subsection{Flux-flux analysis}

We perform a flux-flux analysis to decompose the observed light curves into 
variable and constant components, following the approach of \citet{McHardy2018, JVHS_F9_2020, Cackett20} and similar to the method of  \citet{Winkler1997}, extended here to a simultaneous multi-band fit, using the analysis tools from \cite{Miller2023}.
The light curves are modelled as
\begin{equation}
    f_{\nu}(\lambda, t) = A_{\nu}(\lambda) + R_{\nu}(\lambda)\,X(t)\ ,
\label{eq:flux_flux}
\end{equation}
where $A_{\nu}(\lambda)$ is the constant component, $R_{\nu}(\lambda)$ 
is the rms spectrum of the variable component, and $X(t)$ is the 
normalized driving light curve with zero mean and unit standard deviation. Accordingly, the flux--flux decomposition is carried out over the $ugriz$ bands.

\begin{figure*}
\centering
\includegraphics[width=1\linewidth]{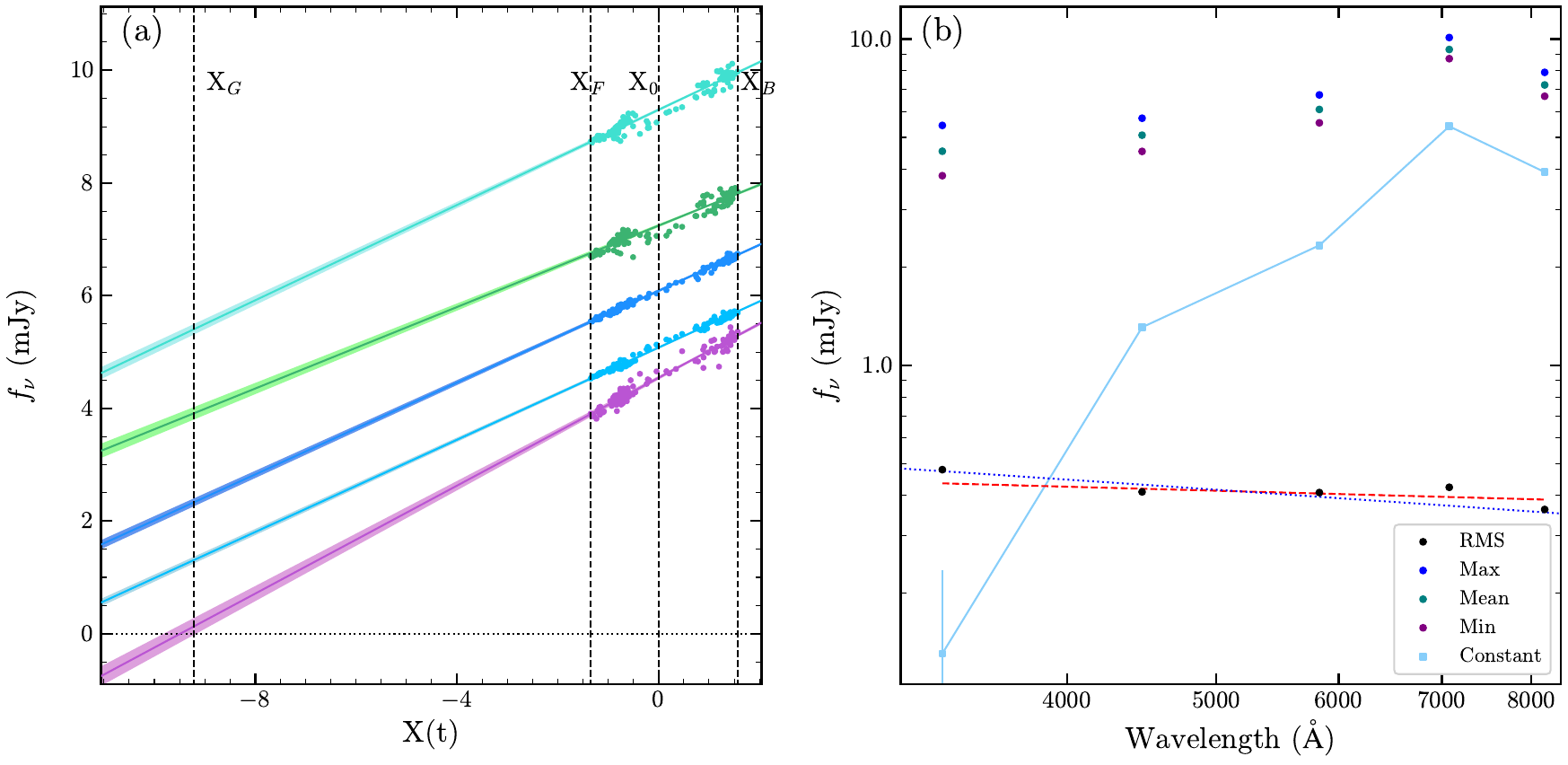}
\caption{Flux–flux decomposition for Mrk 304. Panel (a) illustrates the linear relationship between flux and the model light curve $\mathrm{X}(t)$ for each band. The lines show the fit of Eq. \ref{eq:flux_flux} to each band and its 1 $\sigma$ uncertainty envelope. Here, $\mathrm{X}_G$ denotes the flux value adopted to estimate the constant flux component in each band. The slope of the fitted line for a given filter provides the variability amplitude parameter, $R_{\nu}(\lambda)$. In addition, $\mathrm{X}_F$ and $\mathrm{X}_B$ represent the faintest and brightest flux states observed in the light curves, respectively, whereas $\mathrm{X}_0$ corresponds to the mean flux level. Panel (b) shows the fluxes in each band as a function of wavelength. Overlaid are power-law models for a standard $\lambda^{-1/3}$ disk (dotted blue) fitted to the rms spectrum.
A free-parameter fit (dashed red) yields a slope of $-0.13 \pm 0.11$. During the analysis, we used $E(B-V)=0.073 $ \citep{SFD98}.
}
\label{fig:fluxflux}
\end{figure*}

The results are presented in Fig.\,\ref{fig:fluxflux} and 
Table\,\ref{tab:flux_flux} in the rest-frame assuming Galactic $E(B-V)=0.073 $ \citep{SFD98}. Panel (a) plots 
the band fluxes (in mJy) against $X(t)$; the $y$-intercept of each linear relation gives the constant contribution $A_{\nu}(\lambda)$, while the slope gives the rms variability amplitude $R_{\nu}(\lambda)$. As expected, longer wavelength bands show a larger constant contribution, as expected for an increasingly dominant host galaxy at redder wavelengths (see Table\,\ref{tab:flux_flux}). Panel (b) plots $R_{\nu}(\lambda)$ against effective wavelength. Assuming that the variable component is a standard Shakura-Sunyaev thin disk \citep{SS73}, then it is expected to follow $R_{\nu} \propto \lambda^{-1/3}$. We fit the rms spectrum 
with two models: fixing the power-law index to $-1/3$ (thin disk; dotted blue line) and 
allowing it to vary freely (dashed red line). 
The free fit yields a best-fit index of 
$\beta = -0.13 \pm 0.11$. We note that both the $u$ and $i$ bands show an excess rms flux, consistent with contributions from the BLR diffuse continuum ($u$ band) and a strongly variable H$\alpha$ line ($i$ band).

\begin{deluxetable*}{lccccc}
\centering
\tablecaption{Fluxes (mJy) determined from the flux-flux analysis. \label{tab:flux_flux}}
\tablehead{\colhead{Parameter} & \colhead{u} & \colhead{g} & \colhead{r} & \colhead{i} & \colhead{z} }
\startdata
Max & $5.4\pm 0.1$ & $5.7\pm 0.03$ & $6.8\pm 0.1$ & $10.1 \pm 0.1$ & $7.9 \pm 0.1$  \\
Mean & $4.5 \pm 0.01$ & $5.1\pm 0.01$ & $6.1 \pm 0.01$ & $9.3\pm 0.01$ & $7.2 \pm 0.01$  \\
Min & $3.8 \pm 0.1$ & $4.5 \pm 0.02$ & $5.5 \pm 0.02$ & $8.7 \pm 0.04$ & $6.7 \pm 0.1$  \\
Constant & $0.13\pm0.10$ & $1.3\pm0.03$ & $2.3\pm0.04$ & $5.4\pm0.1$ & $3.9\pm0.1$  \\
RMS & $0.5 \pm 0.01$ & $0.4 \pm 0.01$ & $0.41 \pm 0.01$ & $0.4 \pm 0.01$ & $0.4 \pm 0.01$  
\enddata
\end{deluxetable*}

\section{Discussion}\label{sec:disc}

We monitored the X-ray weak AGN Mrk 304 for $\sim$5.5 months at near-daily cadence in the $ugriz$ photometric bands.
We present the results of our time series analysis of these data. Inter-band time lags, measured consistently with both \texttt{PyCCF} and \texttt{JAVELIN}, confirm that variations in the bluer bands precede those in the redder ones. The resulting lag-wavelength spectrum is broadly consistent with the standard thin-disk prediction ($\tau \propto \lambda^{4/3}$), but a significant excess lag is robustly detected in both the $u$ and $i$ bands, deviating from this simple model. A flux-flux analysis of the $ugriz$ bands allowed us to decompose the light curves into constant and variable components. The rms spectrum is broadly consistent with an accretion disk with $R_{\nu} \propto \lambda^{-1/3}$.
In addition to our high cadence campaign, we analyzed 8 years of ZTF $g$ and $r$ monitoring of Mrk 304. Analysis of the ZTF data reveals a stable $g-r$ lag between 2018 and 2025, with little annual variation.  

In both the lags and the rms spectra, the $i$ band sits above the wavelength-dependent trend. This deviation is consistent with strong line emission features in the filter window.

Our results for Mrk 304 align with broader literature trends. Specifically, the large $\tau_0$ value found in our model is comparable to those reported by, e.g., \cite{Miller2025, Guo2022_ZTF_arxiv, Steyn26} for substantially different samples. 
We note how the lag-wavelength relation from the lowest-frequency bin of our lag-frequency analysis is closer to the SSD prediction. However, as we discuss below, in the specific case of Mrk 304, multiple observables point to the reverberation signal being dominated by non-disk emission.

\subsection{Evidence for non-disk reprocessing}

We summarize the evidence for a non-disk reverberation signal below:

\begin{itemize}
    \item The measured $\tau_0$ is substantially larger than the simplest analytical disk size estimates.
    \item There is an observed $u$-band excess lag, suggesting substantial emission from the BLR diffuse continuum.
    \item The lag-frequency analysis shows that the reverberation signal comes from only the longest timescales probed by the light curve, indicating reprocessing on sizescales greater than about 30 light days.
    \item The excess lag in the $i$ band is likely due to strong H$\alpha$ emission.
\end{itemize}

Put together, this paints a picture of reverberation from beyond the accretion disk, presumably the BLR or wind. That the lags and rms spectrum broadly follow the wavelength-dependence for a disk is not in itself counter to this - diffuse continuum emission from the BLR is expected across the full wavelength range, with the wavelength dependence being close to the expectations for a disk \citep{KoristaGoad2001}. The observed diffuse continuum lag at a given wavelength is a combination of the emissivity-weighted radius, the diffuse continuum flux fraction and the properties of the driving light curve. For an example and detailed explanation, see the photoionization calculations and simulations in \cite{KoristaGoad2001, Korista2019} to see the expected wavelength-dependent lags from the diffuse continuum.

The lag-frequency analysis shows the lags consistent with zero up to a frequency of approximately 0.03 day${^-1}$.  Given the length of the light curve, the lowest frequency bin is large and contains few Fourier frequencies, but the unbinned lags suggest a steep increase at lower frequencies in the $i$ and $z$ bands.  To obtain more frequency bins and better measure the frequency-dependence of the lags will require monitoring of Mrk~304 in future years. This will then allow for comparison to disk and BLR models.

\subsection{\texttt{PyCCF} vs. \texttt{JAVELIN}}
Having established that the observed lags likely involve components more complex than simple disk reprocessing, we now validate the robustness of our measurements using independent techniques.

We determine the disk lag-wavelength relation $\tau \propto \lambda^\beta$ fit to lags measured using both the 
\texttt{PyCCF} and \texttt{JAVELIN} methods.
The lag normalization is characterised by 
the parameter $\tau_0$, evaluated at the $g$-band reference wavelength ($\lambda_0 = 4770$ \AA).
With the slope fixed at the thin-disk prediction of $\beta = 4/3$, both techniques recover 
comparable normalization parameters: $\tau_0^{\text{CCF}} = 6.6 \pm 0.5$ days and 
$\tau_0^{\text{JAV}} = 5.8 \pm 0.2$ days. The agreement to within $\sim 13\%$ indicates 
robust constraints on the overall lag amplitude, independent of the method to measure the lags.

When allowing the slope to vary freely, the CCF analysis yields $\beta_{\text{CCF}} = 1.4 \pm 0.8$, which remains consistent with the theoretical value of $4/3$. Notably, the CCF normalization $\tau_0^{\text{CCF}} = 6.5 \pm 5.4$ days remains nearly identical to the fixed-slope result, validating that the CCF lags are robustly consistent with a 4/3 slope, albeit with a larger normalization. In contrast, the \texttt{JAVELIN} fit favors a significantly steeper scaling of $\beta_{\text{JAV}} = 2.4 \pm 0.3$. This comparison highlights a distinct slope-normalization trade-off: \texttt{PyCCF} preserves a slope near the theoretical prediction while maintaining a high $\tau_0$, whereas \texttt{JAVELIN} adopts a much steeper wavelength dependence with a significantly reduced $\tau_0$ ($\tau_0^{\text{JAV}} = 2.3 \pm 0.4$ days). This divergence in the free-slope regime reflects differences in statistical weight and wavelength coverage rather than a fundamental disagreement on the physical scale of the reverberation response.

\subsection{\texttt{PyCCF}: ZTF vs. Zowada}

Zowada broadband monitoring in 2025 yields 
lag values of $2.8^{+0.8}_{-0.8}$ days (centroid) and $1.6^{+2.0}_{-1.0}$ 
days (peak) between $g$ and $r$. In contrast, ZTF photometry in 2025 provides only a single band pair, 
$g$-$r$, yielding a centroid lag of $6.0^{+1.9}_{-1.8}$ days and a peak lag 
of $5.5^{+2.0}_{-2.5}$ days. While both the ZTF centroid and peak lags are longer than those from Zowada, the lag uncertainties are also much larger. The differences in the lags are not statistically significant, with the peak lags consistent within 1$\sigma$ and the centroid lags only slightly more than 1$\sigma$ different. In general, we find that the ZTF light curves from other years show a greater consistency with the Zowada lag measurements. However, these larger uncertainties with ZTF are due to the lower cadence and lower S/N flux measurements. Using Zowada, we obtained a cadence $\Delta t = 1.3$ days in the $g$ band, whereas $\Delta t \sim 3$ for ZTF. We achieved average photometric uncertainties of 0.5\% in the $g$ and $r$ bands with Zowada, while the average for ZTF is 1\%.

We explored combining the Zowada and ZTF light curves to assess whether the additional baseline and sampling improve the inter-band lag measurements, particularly for the $g$--$r$ lag.
Prior to any joint analysis, the two datasets were intercalibrated using \texttt{PyCALI} \citep{Li2014}, which models the flux offset and multiplicative scaling between datasets while accounting for the different photometric systems and apertures of the two telescopes. The lag recovered from the combined light curves is consistent with that derived from the Zowada-only dataset within $1\sigma$, confirming that the two datasets describe the same underlying variability signal. However, the inclusion of ZTF data did not improve the lag precision and inflated the uncertainties. The ZTF observations have a sparser cadence compared to Zowada and carry larger photometric uncertainties, and their inclusion introduces additional low-frequency power into the merged light curve that distorts the shape of the underlying variability trend. This effectively degrades the cross-correlation signal rather than reinforcing it. We therefore adopted the Zowada-only light curves for the 2025 lag measurements, treating the combined analysis as a consistency check.

\subsection{X-ray weakness and UV/optical variability}
Mrk~304 presents a compelling case study for understanding accretion physics in X-ray weak AGN, 
where the standard X-ray reverberation paradigm faces its greatest challenges.
The archival \textit{NuSTAR} observations were not taken during the Zowada campaign, but were taken only 7 months before the start.  The broadband \textit{NuSTAR} X-ray spectrum is well described by a power-law continuum with photon index $\Gamma = 1.78\pm0.03$. 
The 2--10~keV flux is $F_{2-10} = (6.9 \pm 0.23)\times10^{-12}\,{\rm erg\,cm^{-2}\,s^{-1}}$, corresponding to a luminosity of $L_{2-10} = (7.6 \pm 0.3)\times10^{43}\,{\rm erg\,s^{-1}}$ reflects the source's characteristically weak X-ray emission. Applying the bolometric correction of \cite{Marconi2004} we estimate $L_{\mathrm{bol}} = (2.3\pm0.1) \times 10^{45} \,{\rm erg\,s^{-1}}$.

To investigate the energy between the corona and the accretion disk, we calculated the broad-band X-ray to optical/UV luminosity ratio. The luminosity of the {\it variable} part of the accretion disk was determined to be $L_{\text{disk}} \approx 1.6 \times 10^{45}$~erg~s$^{-1}$ from fitting the rms variability in the \textit{u, g, r, i,} and \textit{z} bands with the SSD. For the high-energy component, we extrapolated the best-fitting model using \texttt{XSPEC} \texttt{energies} to estimate a total X-ray luminosity of $L_{\text{X}} = 2.9 \times 10^{44}$ erg s$^{-1}$ over the 0.03--200 keV energy range. This results in an X-ray to optical/UV luminosity ratio of $L_{\text{X}} / L_{\text{disk}} \approx 0.19$. This value demonstrates that the total power emitted in X-rays (not just the variable component) is only about 19\% of the variable optical/UV luminosity from the disk. Such a low ratio demonstrates that the X-ray emission is energetically insufficient to be the primary driver of the observed optical/UV variability through simple disk reprocessing models.  

The measured UV/optical lags measured in Mrk 304 show a similar wavelength dependence to many of the other AGN where such continuum RM has been undertaken.  Comparing the magnitude of the lags in Mrk 304 with other AGN, with an approximate $\log L_{5100}$ of 44.4 erg/s, and $\log{\tau_{5100}} = 0.9$ days, Mrk 304 sits very close to the best-fitting lag-luminosity relation from \citet{Miller2025}.  In other words, based on its continuum RM properties only, there is nothing that separates Mrk 304 from other AGN that have been studied.

What is remarkable about Mrk 304, however, is its X-ray weakness.  The presence of measurable inter-band lags in Mrk~304, despite its established X-ray weakness \citep[][and this work]{Brinkmann2004}, presents a challenge to the standard reprocessing scenario where a compact, X-ray-bright corona irradiates and drives the correlated UV/optical variability across the disk.  If X-ray irradiation were the primary driver, one would expect the lag amplitude and coherence to scale with X-ray flux; the detection of significant lags in an X-ray weak source instead implies that the X-ray corona is not the dominant source of variability input, or at least not the sole one.

We can also compare our result to simulations. For instance, MHD disk simulations by \citet{Secunda2024, Secunda2025} conclude that X-rays can drive reverberation, but only if they are luminous enough and the disk albedo is low enough. \citet{Secunda2025} found a threshold of $L_{\text{X}} / L_{\text{disk}} < 0.25$ below which X-rays are not strong enough to drive reverberation, which aligns well with our value of 0.19, reinforcing that Mrk 304 does not have luminous enough X-rays.  Studying the properties of the multi-year variability of Fairall 9, \citet{Hagen2024} concluded that the majority of the observed variability is driven by intrinsic variability of the disk itself through propagating fluctuations, and that the observed reverberation signal is dominated by the EUV (not the X-ray) illuminating a wind, rather than the disk.  \citet{Hagen2025} took this further, performing Fourier analysis of the lags, showing that the connection between the X-ray and UV is more complicated than the picture in \citet{Hagen2024} and that both the UV and X-ray are intrinsically variable and only weakly correlated.  Moreover, changes in the SED suggest the inner structure is likely changing.  The situation in Mrk~304 may be similar, with the lags dominated by the EUV (which should have the required luminosity) illuminating a wind. Both the energetics and the observed lag properties of Mrk~304 point to this picture. 
In the scenario where the EUV is being reprocessed in the BLR, resulting in the continuum lags, we overcome the problem of X-ray weakness and explain why the frequency-resolved lags suggest reprocessing from beyond the accretion disk.

Alternatively, other mechanisms have been suggested for the origin of continuum lags. For instance, the magnetically coupled corona-disk system described by the CHAR model \citep{MSun2020_CHAR_model} provides an alternative geometry in which magnetic fields threading the corona and disk couple the two components, transferring energy and producing correlated variability without requiring strong X-ray irradiation of the outer disk. Intrinsic temperature fluctuations propagating through the disk \citep{Dexter_Agol2011, Ren2024} offer a further mechanism that naturally produces wavelength-dependent lags consistent with the observed $\tau \propto \lambda^{4/3}$ scaling, but without invoking a centrally irradiating source.  

In summary, the presence of well-defined continuum lags, occurring on timescales $>$30 days, in the X-ray weak AGN Mrk~304 challenges the standard reprocessing picture where X-rays drive variability in the disk at longer wavelengths.  This suggests either the reverberation lags have a different origin, for instance, EUV radiation illuminating a wind \citep{Hagen2024}, or that the lags are not reverberation \citep[e.g.,][]{MSun2020_CHAR_model}.  Further monitoring of X-ray weak AGN, particularly with simultaneous X-ray monitoring, can help shed light on the mechanisms involved.

\section*{Acknowledgements}
We thank the anonymous referee for their detailed and thoughtful comments that have improved this manuscript. 

\appendix
\restartappendixnumbering

\section{Accretion rate and bolometric luminosity determination}
We determined bolometric luminosity and accretion rate using different methodologies, i.e., \cite{Marconi2004, Netzer2019}, utilizing the optical and X-ray fluxes, finding general consistency between the two approaches.

\begin{deluxetable}{lccc}[ht]
\centering
\tablecaption{Bolometric luminosity, Eddington ratio, and mass accretion rate estimates using different methods. \\$L_{\rm Edd} = 2.5 \times 10^{46}$ [erg s$^{-1}$]; $\lambda_{\rm Edd}$ = $L_{\rm bol}$/ $L_{\rm Edd}$. \label{tab:bolometric_estimates}}
\tablehead{
\colhead{Method} & \colhead{$L_{\rm bol}$ [erg s$^{-1}$]} & \colhead{$\lambda_{\rm Edd}$} & \colhead{$\dot{M}$ [$M_\odot$ yr$^{-1}$]}}
\startdata
Optical; (fixed BC = 10.3)   & $2.9 \times 10^{45}$ & 0.1 & 0.5 \\
Optical; \cite{Netzer2019} $L_{5100}$ & $3.6 \times 10^{45}$ & 0.1 & 0.6 \\
X-ray; \cite{Netzer2019} $L_{2-10\,\mathrm{keV}}$ & $1.9 \times 10^{45}$ & 0.1 & 0.3 \\
X-ray; \cite{Marconi2004}        & $2.3 \times 10^{45}$ & 0.1 & 0.4
\enddata
\end{deluxetable}

\section{ZTF light curves}

Properties of the ZTF light curves are given in Table.~\ref{tab:season_properties}.  The light curves from each year, along with the CCF and lag distributions, are shown in Fig.~\ref{fig:mrk304}.

\begin{figure*}
\centering
\includegraphics[width=1\textwidth]{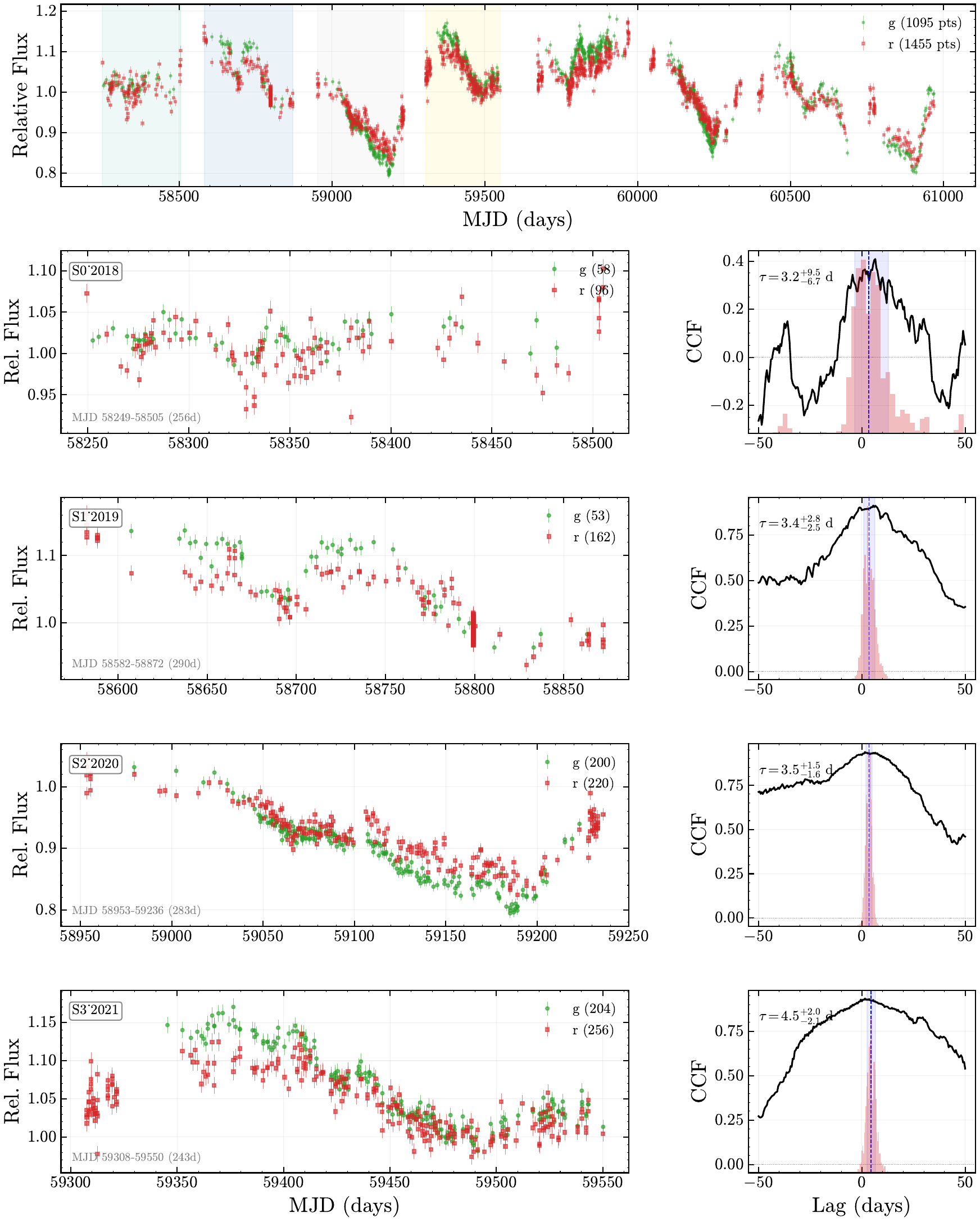}

\caption{Light curves of Mrk 304 and lag distribution from ZTF. The top panel is the summary of all of the years' analysis. Divided by years according to gaps, each year has a different shaded color. The left panel has $g$ in green and $r$ in red. The right panel shows the lag distribution measured by \texttt{PyCCF}.}
\label{fig:mrk304}
\end{figure*}

\begin{figure*}
\ContinuedFloat
\centering
\includegraphics[width=1\textwidth]{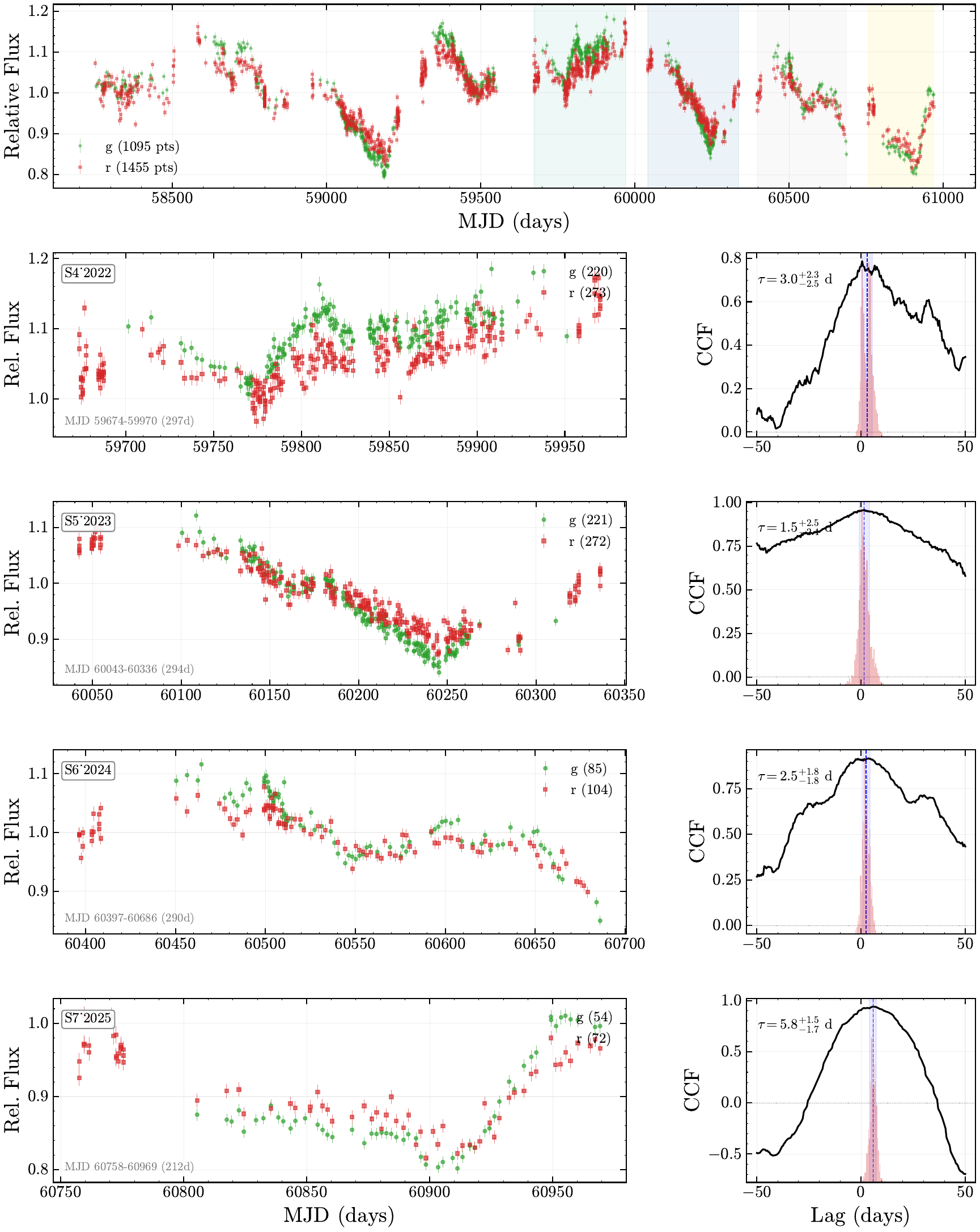}

\caption{Continued.}
\end{figure*}

\begin{deluxetable*}{llccccccc}
\centering
\tablecaption{Seasonal light curve properties for ZTF data.\label{tab:season_properties}}
\startdata
\tablehead{
\colhead{Year} & \colhead{Band} & \colhead{N} & \colhead{Start} & \colhead{End} & \colhead{Length [d]} & \colhead{$\langle \Delta t \rangle$} & \colhead{Median $\Delta t$} & \colhead{$F_{\rm var}$} \\
}2018 & g & 58 & 58252.5 & 58482.1 & 229.6 & 4.0 & 3.0 & 0.01 \\
2018 & r & 96 & 58249.5 & 58505.1 & 255.6 & 2.7 & 1.0 & 0.03 \\
2019 & g & 53 & 58607.5 & 58863.1 & 255.6 & 4.9 & 3.0 & 0.05 \\
2019 & r & 162 & 58582.5 & 58872.1 & 289.6 & 1.8 & 0.0 & 0.04 \\
2020 & g & 200 & 58979.4 & 59223.1 & 243.7 & 1.2 & 1.0 & 0.05 \\
2020 & r & 218 & 58953.5 & 59236.1 & 282.6 & 1.3 & 1.0 & 0.04 \\
2021 & g & 204 & 59345.4 & 59550.2 & 204.7 & 1.0 & 0.9 & 0.05 \\
2021 & r & 256 & 59307.5 & 59550.1 & 242.6 & 1.0 & 0.1 & 0.03 \\
2021 & g & 204 & 59345.4 & 59550.2 & 204.7 & 1.0 & 0.9 & 0.05 \\
2021 & r & 256 & 59307.5 & 59550.1 & 242.6 & 1.0 & 0.1 & 0.03 \\
2022 & g & 220 & 59701.5 & 59951.1 & 249.6 & 1.1 & 0.8 & 0.03 \\
2022 & r & 273 & 59673.5 & 59970.1 & 296.6 & 1.1 & 0.2 & 0.03 \\
2023 & g & 221 & 60100.5 & 60311.1 & 210.7 & 1.0 & 0.2 & 0.07 \\
2023 & r & 271 & 60042.5 & 60336.1 & 293.6 & 1.1 & 0.1 & 0.06 \\
2024 & g & 85 & 60450.4 & 60686.1 & 235.7 & 2.8 & 2.0 & 0.05 \\
2024 & r & 104 & 60396.5 & 60679.2 & 282.6 & 2.7 & 2.0 & 0.04 \\
2025 & g & 54 & 60805.5 & 60969.2 & 163.8 & 3.1 & 2.1 & 0.07 \\
2025 & r & 72 & 60757.5 & 60969.3 & 211.7 & 3.0 & 2.0 & 0.06 \\
\enddata
\end{deluxetable*}


\section{Kernal Selection}
There are common forms for the kernel function as those assessed to describe AGN variability, such as Radial Basis Function (RBF) Rational Quadratic (RQ) and Matérn kernels. Following, e.g., \cite{Wilkins2019, Lewin2023}, we test RBF, Matérn 1/2, Matérn 3/2, Matérn 5/2, RQ, and auto options that Stela allows. 
Lags are statistically consistent and do not depend on the choice of the kernel. We have found that Akaike Information Criterion (AIC) and Bayesian Information Criterion (BIC) tests are identical, $\pm$ 0.3 in AIC and $\pm$ 0.1 BIC, but rational quadratic (RQ). AIC 4.2-4.5 and BIC 12.87-18.89 were for all but RQ 6.26 and 17.76, respectively, for the g-band. A lower AIC or BIC indicates that a model is preferred over the others being compared. We have chosen Matérn 3/2 as the best kernel. For more information about functional forms, please see \cite{Wilkins2019}.

\begin{figure*}
\centering
\includegraphics[width=1\textwidth]{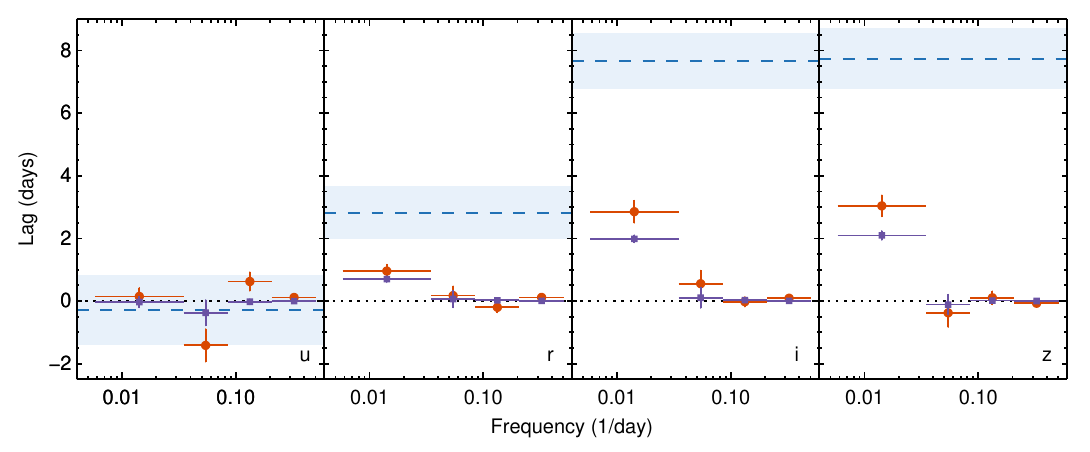}
\caption{The lag comparison with respect to the $g$ band, as a function of frequency for each of the $u, r, i$ and $z$ light curves, for linearly interpolated light curves onto a regular 1-day grid (orange circles) and STELA's with Matérn 3/2 kernel (purple squares). The blue dashed line and region show the CCF centroid lags.}
\label{fig:linear_vs_stela}
\end{figure*}

\bibliography{bibliography}{}
\bibliographystyle{aasjournal}

\end{document}